\documentclass[
  aip,            
  reprint,        
  floatfix,       
  10pt
]{revtex4-2}
\usepackage[utf8]{inputenc}
\usepackage[T1]{fontenc}
\usepackage{amssymb}
\usepackage{amsmath}
\usepackage{booktabs}
\usepackage{physics}
\usepackage{graphicx}
\usepackage{bibentry}
\usepackage{tikz}
\usepackage{curves}
\usepackage{multirow}
\usepackage{soul}
\usepackage{placeins}

\usepackage{xcolor}

\definecolor{RoyalBlue}{cmyk}{1, 0.50, 0, 0}
\definecolor{ForestGreen}{rgb}{0, 0.608, 0.333}
\definecolor{RawSienna}{rgb}{0.59, 0.25, 0.02}

\allowdisplaybreaks

\newcommand{\Exp}[1]{\mathrm{e}^{#1}}

\newcommand{\rewop}[1]{_{\mathrm{#1}}}

\newcommand{\BesselJ}[2]{\mathrm{J}_{#1} {\pqty{#2}}}

\newcommand{\EE}[1]{E\rewop{#1}}
\newcommand{\VV}[1]{V\rewop{#1}}
\newcommand{\Sec}[1]{Sec.~\ref{sec:#1}}
\newcommand{\Fig}[1]{Fig.~\ref{fig:#1}}

\begin{document}
\title{A passive atomtronics filter for Fermi gases}
\author{Jun Hao Hue}
\email{jhhue@nus.edu.sg}
\affiliation{Centre for Quantum Technologies, National University of Singapore, 3 Science Drive 2, Singapore 117543}
\author{Martin-Isbj\"{o}rn Trappe}
\email{martin.trappe@quantumlah.org}
\affiliation{Centre for Quantum Technologies, National University of Singapore, 3 Science Drive 2, Singapore 117543}
\author{Piotr T. Grochowski}
\email{piotr.grochowski@upol.cz}
\affiliation{Department of Optics, Palack\'y University, 17.~listopadu 1192/12, 771 46 Olomouc, Czech Republic}
\author{Jonathan Lau}
\email{jonathanlau66@gmail.com}
\affiliation{Centre for Quantum Technologies, National University of Singapore, 3 Science Drive 2, Singapore 117543}
\author{Leong-Chuan Kwek}
\email{cqtklc@nus.edu.sg}
\affiliation{Centre for Quantum Technologies, National University of Singapore, 3 Science Drive 2, Singapore 117543}
\affiliation{National Institute of Education,
Nanyang Technological University, 1 Nanyang Walk, Singapore 637616}
\affiliation{School of Electrical and Electronic Engineering, Nanyang Technological University, Block S2.1, 50 Nanyang Avenue, Singapore 639798}
\affiliation{Quantum Science and Engineering Center, S1-B4a-02, 50 Nanyang Avenue, Singapore 639798}
\affiliation{MajuLab, CNRS-UNS-NUS-NTU International Joint Research Unit, UMI 3654, Singapore 117543}
\date{\today}

\begin{abstract}
We design an atomtronic filter device that spatially separates the components of a two-component Fermi gas with repulsive contact interactions in a two-dimensional geometry. With the aid of density--potential functional theory (DPFT), which can accurately simulate Fermi gases in realistic settings, we propose and characterize a barbell-shaped trapping potential, where a bridge-shaped potential connects two ring-shaped potentials. In the strongly repulsive regime, each of the ring traps eventually stores one of the fermion species. Our simulations are a guide to designing component filters for initially mixed, weakly repulsive spin components. We demonstrate that the functioning of this barbell design is robust against variations in experimental settings, for example, across particle numbers, for small deformations of the trap geometry, or if interatomic interactions differ from the bare contact repulsion. Our investigation marks the first step in establishing DPFT as a comprehensive simulation framework for fermionic atomtronics.
\end{abstract}

\maketitle

\section{Introduction}

Atomtronics is an emerging quantum technology that promises advances in quantum simulations, matter wave computing circuits, and quantum metrology by harnessing the unique properties of ultracold atomic gases in lieu of electronic circuit elements. While several traditional electronic components, such as batteries, transistors, and diodes, have been recreated as atomtronic elements \cite{Seaman_2007, Stickney_2007, Pepino_2009, Pepino_2021}, the inherently quantum-mechanical nature of matter wave circuits constructed from ultracold quantum gases can also manifest capabilities that do not have analogs in traditional electronics
\cite{amico2022colloquium,amico2021roadmap}. One example is the Datta-Das transistor \cite{vaishnav2008spin}. Such unconventional atomtronic circuit components are designed, for instance, by manipulating the quantum statistics of trapped atoms \cite{haug2019andreev,haug2019aharonov} or the geometry of the trapping potentials \cite{Henderson2009}. An atomtronic analog of a superconducting quantum interference device (SQUID) can also be constructed with ultracold bosonic atoms that are confined to a ring-shaped potential with a barrier in the ring \cite{Fagaly_2006,Ramanathan_2011, Eckel_2014}. Similarly to the bosonic case, transversal magnetic fields can induce persistent currents in ultracold Fermi gases \cite{Wright_2013, Cai_2022}, potentially enriching the toolbox for quantum simulations \cite{beregi2024quantum} and quantum sensing \cite{Helm2015, amico2021roadmap, amico2022colloquium, Pezze2024}. However, theoretical studies on the behavior of such Fermi gases in atomtronic elements tend to focus on idealized settings with small particle numbers. For example, two-dimensional ring-shaped potentials are modeled by one-dimensional lattices with periodic boundary conditions, and the few-body system is then solved using exact diagonalization or density matrix renormalization group methods \cite{Chetcuti2022, Osterloh2023}. Recently, time-dependent density functional theory predicted persistent currents of superfluid fermions in a ring-shaped potential \cite{Xhani2024}, albeit for particle numbers smaller than typical experiments demand.

In this work, we realize a passive fermion filter and calculate accurate density distributions of mesoscopic ultracold atomic Fermi gases in tailored two-dimensional traps, thereby narrowing the gap between (i) the existing small-scale simulations of fermion-based atomtronic building blocks and (ii) the required large-scale designs for realistic experimental conditions. We extract the ground-state density distributions using density--potential functional theory (DPFT) \cite{Englert1988,Trappe2016,Trappe2017,Englert2019,Englert2023DFMPS}, an orbital-free variant of density functional theory \cite{Ligneres2005,Chen2008,Burke2012,Witt2018}, which has an established track record across the physical sciences \cite{Trappe2019,Trappe2023NatComm,Trappe2023_nanoparticle}. DPFT is particularly suitable for simulating ultracold Fermi gases in low-dimensional traps \cite{Trappe2016,Trappe2017,Chau2018,Trappe2021b,Trappe2023DFMPS,Grochowski2024}, where even the basic DPFT approximations make predictions of a quality comparable to Hartree--Fock theory \cite{Trappe2021b}. The resulting DPFT density profiles of the fermionic clouds guide our trap designs. First for simulate Fermi gases with small particle numbers, where a number of intricate density separation patterns are most clearly visible. Then, we move to a realistic setup, which aligns with recent experiments \cite{Cai_2022}, and determine quantitatively how the two spin-components of a gas of 20000 Lithium-6 atoms can be separated in our newly designed atomtronic spin filter.

This work is organized as follows. In \Sec{DPFT}, we briefly describe the concepts behind DPFT, the systematic DPFT approximations that feature in this work, and the main characteristics of the two-component contact-interacting Fermi gases as prescribed by experimental setups. In \Sec{Filter}, we develop the blueprint of the atomtronic filter component, which we coin `barbell' potential due to its shape. Section~\ref{sec:Results} collects our results: In \Sec{10+10}, we analyze the behavior of Fermi gases, with ten contact-interacting particles in each (spin-)component, trapped in various forms of a barbell potential. We thereby design an external trapping potential that facilitates spatial separation of the two Fermi gas components and, hence, acts as a passive atomtronic component filter. These investigations also prepare our computational infrastructure for simulating Fermi gases under actual experimental conditions. Finally, in \Sec{realistic}, we (i) replace the bare contact interaction with a more realistic dressed contact interaction based on quantum Monte Carlo simulations \cite{Grochowski2017,Trappe2021b}, (ii) extend our analysis to large particle numbers, and (iii) simulate a setup aligned with recent experiments \cite{Cai_2022}.

\section{Computational framework and interaction functionals} \label{sec:DPFT}

In the following, we outline the DPFT framework and reiterate the expressions used in this work; for details, see \cite{Chau2018,Englert2019,Trappe2021b,Trappe2023DFMPS,Trappe2023_nanoparticle} and references therein. While the energy functional $\EE{}$ in traditional variants of orbital-free density functional theory (DFT) \cite{Hohenberg_1964,Levy1979} depends only on the particle density $n$ (and the chemical potential $\mu$), the fundamental variables of DPFT are both $n$ and an effective potential $V$ that includes the interaction effects. This bifunctional formulation of the total energy
\begin{equation}\label{eq:Energy}
\begin{aligned}
\EE{}{\bqty{V,n,\mu}} ={}& \EE{1} {\bqty{V - \mu}} - \int {\pqty{\dd{\vb{r}}}} \,{\pqty{V {\pqty{\vb{r}}} - \VV{ext}}}\, n {\pqty{\vb{r}}}\\ 
{}& + \EE{int} {\bqty{n}} + \mu\,N
\end{aligned}
\end{equation}
stems from Legendre-transforming the kinetic energy density functional $\EE{kin}  {\bqty{n}}$ by introducing the new variable ${V{\pqty{\vb{r}}}=\mu-\frac{\delta \EE{kin} {\bqty{n}}}{\delta n{\pqty{\vb{r}}}}}$ at spatial position $\vb{r}$, i.e.,
\begin{equation}
    \EE{1} {\bqty{V - \mu}} = \EE{kin} {\bqty{n}} + \int {\pqty{\dd{\vb{r}}}} \, {\pqty{V {\pqty{\vb{r}}} - \mu}}\, n {\pqty{\vb{r}}}\, .
\end{equation}

In the case of two-component fermions (e.g., spin-1/2, with labels `$\pm$'), the variational equations from the functional derivatives of $\EE{}{\bqty{V,n,\mu}}$ for self-consistently determining the ground-state density read
\begin{align} \label{eq:SelfConsist}
n^\pm {\pqty{\vb{r}}} &= \fdv{\EE{1} {\bqty{V^\pm - \mu^\pm}}}{V^\pm {\pqty{\vb{r}}}} \\
\intertext{and}
V^\pm {\pqty{\vb{r}}} &= \VV{ext}^\pm {\pqty{\vb{r}}} + \fdv{\EE{int} {\bqty{n^+, n^-}} }{n^\pm {\pqty{\vb{r}}}} \,,
\end{align}
with the external potentials $\VV{ext}^\pm$ for the two fermion components, and the particle numbers. Here, $\EE{int} {\bqty{n^+, n^-}}$ is the interaction energy and ${N^\pm=\int {\pqty{\dd{\vb{r}}}} \, n^\pm {\pqty{\vb{r}}}}$ are the particle numbers of each component, which constrain the spatial density functions $n^+{\pqty{\vb{r}}}$ and $n^-{\pqty{\vb{r}}}$. More details on multi-component DPFT can be found in \cite{Trappe2021b}.

In this work we focus on calculating ground-state densities of two-component contact-interacting Fermi gases, for which we use an approximation of the right-hand side of \eqref{eq:SelfConsist} that has been validated for precisely such Fermi gases \cite{Chau2018,Trappe2021b}:
\begin{equation}\label{eq:n3'}
	n_{3'}^\pm {\pqty{\vb{r}}} = \int {\pqty{\dd{\vb{r}'}}} \,{\pqty{\frac{k^\pm_{3'} {\pqty{\vb{r}, \vb{r}'}} }{2 \pi r'}}}^D \BesselJ{D}{2 r' k^\pm_{3'} {\pqty{\vb{r}, \vb{r}'}} } \,,
\end{equation}
where $D$ is the dimension of the system, $\BesselJ{D}{\,}$ is the Bessel function of order $D$, ${r' = \abs{\vb{r}'}}$, 
\begin{equation}
	k^\pm_{3'} {\pqty{\vb{r}, \vb{r}'}} = \frac{1}{\hbar} {\bqty{2 m {\pqty{\mu^\pm - V^\pm {\pqty{\vb{r} + \vb{r}'}}}}}}_{+}^{1/2} \,,
\end{equation}
and ${{\bqty{x}}_+ = x\, \eta {\pqty{x}}}$ with the step function $\eta {\pqty{\,}}$. The expression \eqref{eq:n3'} improves upon the quasiclassical Thomas--Fermi (TF) density by systematically including nonlocal quantum corrections that are crucial for reliable simulations of contact-interacting two-component Fermi gases \cite{trappe2016ground,Trappe2021b}. The kinetic energy consistent with the approximation \eqref{eq:n3'} is
\begin{equation} \label{eq:E3'}
    \begin{aligned}
	\EE{kin,3'}^{\pm} ={}& \frac{ \Omega_D}{{\pqty{2 \pi \hbar}}^D {\pqty{2D + 4}} m } \\
        {}& \times \int {\pqty{\dd{\vb{r}}}}\, {\bqty{2m {\pqty{\mu^\pm - V^\pm {\pqty{\vb{r}}}}} }}_{+}^{\frac{D+2}{2}} \,,       
    \end{aligned}
\end{equation}
where $\Omega_D$ is the solid angle in $D$ dimensions.

In the following, we use harmonic oscillator units, ${\mathcal{E} = \hbar \omega}$ for energy and ${\mathcal{L} = \sqrt{\hbar/(m \omega)}}$ for length. For example, by demanding ${\mathcal{L} =1\,\mu m}$ and taking the mass of lithium-6 atoms, we find $\omega$, which fixes $\mathcal{E}$. We consider Fermi gases that are tightly confined in one spatial dimension, such that we may calculate quantities for strictly ${D = 2}$. We consider the bare contact interaction
\begin{equation} \label{eq:Contact}
	\EE{int} {\bqty{n^+, n^-}} = c \int {\pqty{\dd{\vb{r}}}}\, n^+ {\pqty{\vb{r}}} n^- {\pqty{\vb{r}}}\,,
\end{equation}
with repulsion of strength ${c>0}$ between the fermion components, which is commonly realized experimentally through Feshbach resonances \cite{Chin_2010}. As a more realistic alternative to---and as a comparison with---the bare contact interaction, we also consider a dressed contact interaction that effectively accounts for the finite range of interatomic forces \cite{Grochowski2017,Trappe2021b}:
\begingroup
\setlength{\abovedisplayskip}{10pt}
\setlength{\belowdisplayskip}{10pt}
\begin{equation} \label{eq:Dressed}
\begin{aligned}
    \EE{int} {\bqty{n^+, n^-}} &= \frac{\pi \hbar^2}{m} \int {\pqty{\dd{\vb{r}}}}  n^+ {\pqty{\vb{r}}} n^- {\pqty{\vb{r}}} \\
    &\qquad\times {\bqty{ \beta{\pqty{\eta^+ {\pqty{\vb{r}}}}} + \beta{\pqty{\eta^- {\pqty{\vb{r}}}}} }} \,,
\end{aligned}
\end{equation}
\endgroup
where
\begin{equation} \label{eq:eta}
    \eta^\pm {\pqty{\vb{r}}} = - \frac{1}{\log(k_F^\pm{\pqty{\vb{r}}} a_{\mathrm{2D}})} \,,
\end{equation}
${k_F^\pm = \sqrt{2 \pi n^\pm {\pqty{\vb{r}}}}}$ is the Fermi wave number, and $a_{\mathrm{2D}}$ is the scattering length in two dimensions. The parameter $\beta$ is obtained using the CASINO package, employing the smooth pseudopotential of Whitehead et al. \cite{whitehead2016pseudopotential} to represent interspecies fermionic interactions. Finite-size effects are mitigated by adopting a ${49 + 49}$ closed-shell configuration, and the resulting dependence is expressed through the polynomial fit
\begin{widetext}
\begin{equation} \label{eq:beta}
    \beta {\pqty{\eta}} = \begin{cases}
    \begin{aligned}1.4436 - 0.49271\,\eta^{-1} - 0.036826\,\eta^{-2} - 0.00091746\,\eta^{-3} \end{aligned} \,, & \eta \leq -0.15\\
		\begin{aligned}-0.061398\, \eta^6 + 0.25332\, \eta^5 - 0.30739\, \eta^4 - 0.058454\, \eta^2 + 1.0062\, \eta - 0.00041475\, \end{aligned} \,, & 0 \leq \eta \leq 1.55\\
	1.4436 - 0.46338\, \eta^{-1} - 0.20465\, \eta^{-2} \,, & \eta > 1.55
    \end{cases} \,.
\end{equation}
\end{widetext}

\section{Blueprint of a Component Filter} \label{sec:Filter}

We begin with an analysis of fermions trapped in the ring-shaped potential
\begin{equation} \label{eq:Gaussian}
	V_{R} \pqty{\vb{r}} = - V_0 \;\mathrm{exp}\hspace{-0.5ex}\left[- {\pqty{\frac{2}{\Delta_R}}}^4 {\pqty{r^2 -R^2}}^2\right]
\end{equation}
as the fundamental building block for designing an atomtronic component filter. Here, $R$ is the radius of the ring, ${r = \abs{\vb{r}}}$, $V_0$ is the depth of the potential, and $\Delta_R$ determines the width of the ring. Such ring-shaped traps have been realised experimentally \cite{Cai_2022}. As discussed in the appendix, slightly different functional forms do not alter the targeted qualitative behaviour of the atomtronic device.

\begin{figure}[ht!]
    \centering
    \includegraphics[width=0.9\linewidth]{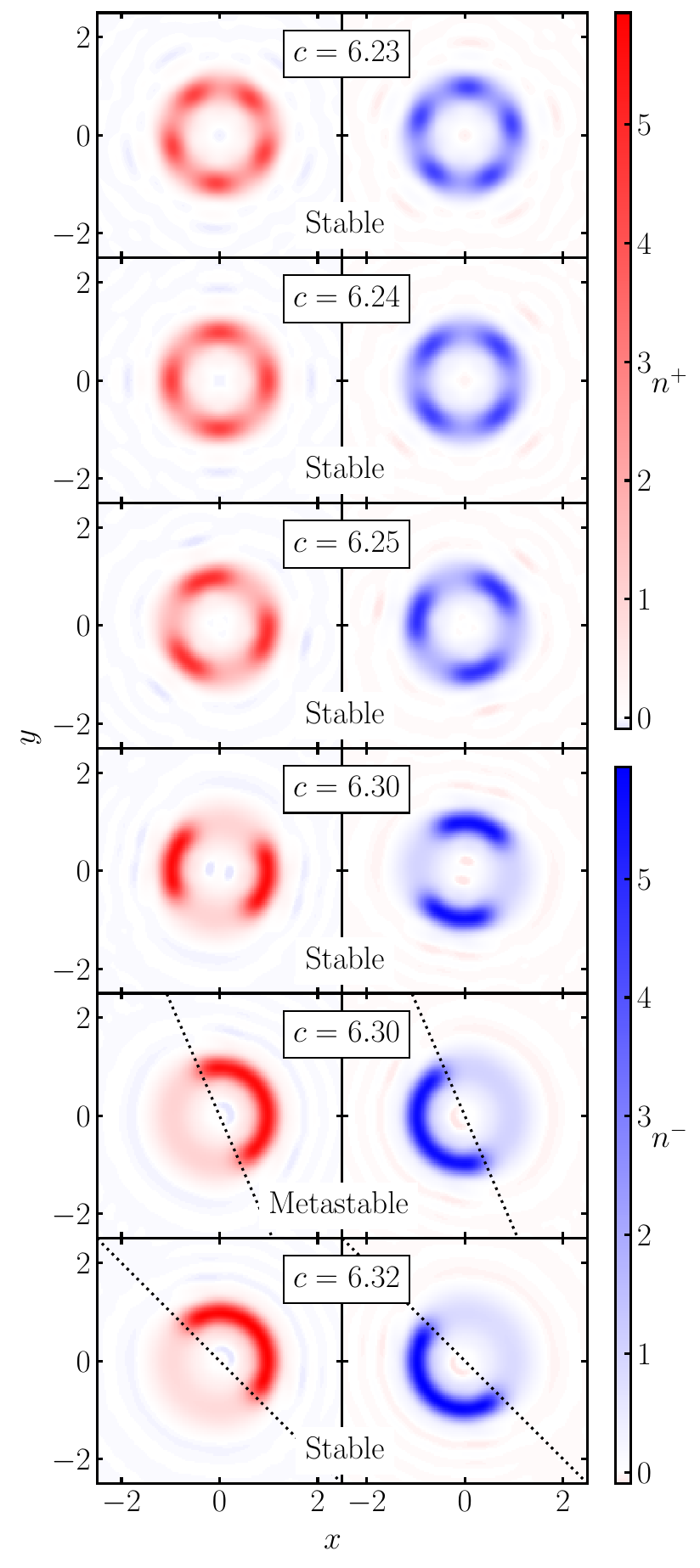}
    \caption{Illustration of the phase transition of a two-component Fermi gas in the ring-shaped potential $V_R$ of \eqref{eq:Gaussian}, with default parameters from Table~\ref{tab:default}, driven by the repulsive contact interaction strength $c$. First separations (into ten domains) are visible at ${c \approx 6.2}$, and the number of domains reduces step-wise until the two components are segregated into one domain each at ${c_{\mathrm{split}} \approx 6.31}$. Due to the rotational symmetry of the system, the domain interfaces can be oriented in any direction (cf. black dashed lines). The observation of the metastable configuration at ${c = 6.3}$, with energy ${E=-1422.037\,\mathcal E}$ (the ground-state energy is ${E=-1422.687\,\mathcal E}$), heralds the transition into the symmetric split as $c$ exceeds $c_{\mathrm{split}}$. We use harmonic oscillator units.}
    \label{fig:Ring}
\end{figure}

We calculate the ground-state particle density profiles $n^\pm{\pqty{\vb{r}}}$, approximated through \eqref{eq:n3'}, within the DPFT framework for ${\VV{ext}^\pm\pqty{\vb{r}}=V_{R} \pqty{\vb{r}}}$. We find (i) the noninteracting state with ${n^+ {\pqty{\vb{r}}} = n^- {\pqty{\vb{r}}}}$ everywhere at repulsion strength ${c=0}$ and (ii) a complete (symmetric) split of the interacting gas as ${c\to\infty}$ beyond a critical $c_{\mathrm{split}}$. In \Fig{Ring} we illustrate the phase transition between both extremes, from a mixed phase at small $c$ (with almost complete overlap, i.e., ${n^+ {\pqty{\vb{r}}} \approx n^- {\pqty{\vb{r}}}}$) to symmetrically separated profiles (with small overlap, i.e., a small spatial region as the interface of $n^+{\pqty{\vb{r}}}$ and $n^-{\pqty{\vb{r}}}$) at large ${c>c_{\mathrm{split}}}$. Intermediate repulsion strengths $c$ yield different patterns of partial separations and also permit metastable states with energies close to the ground-state energy, in line with similar investigations for other traps \cite{Trappe2021b}. While a separation at large $c$ is necessary for component filtering, it is not sufficient: In a single ring potential, the a priori unknown orientation of the interface prohibits filtering without additional spin measurements.

\begin{figure}[htb!]
    \centering
    \includegraphics[width = \linewidth]{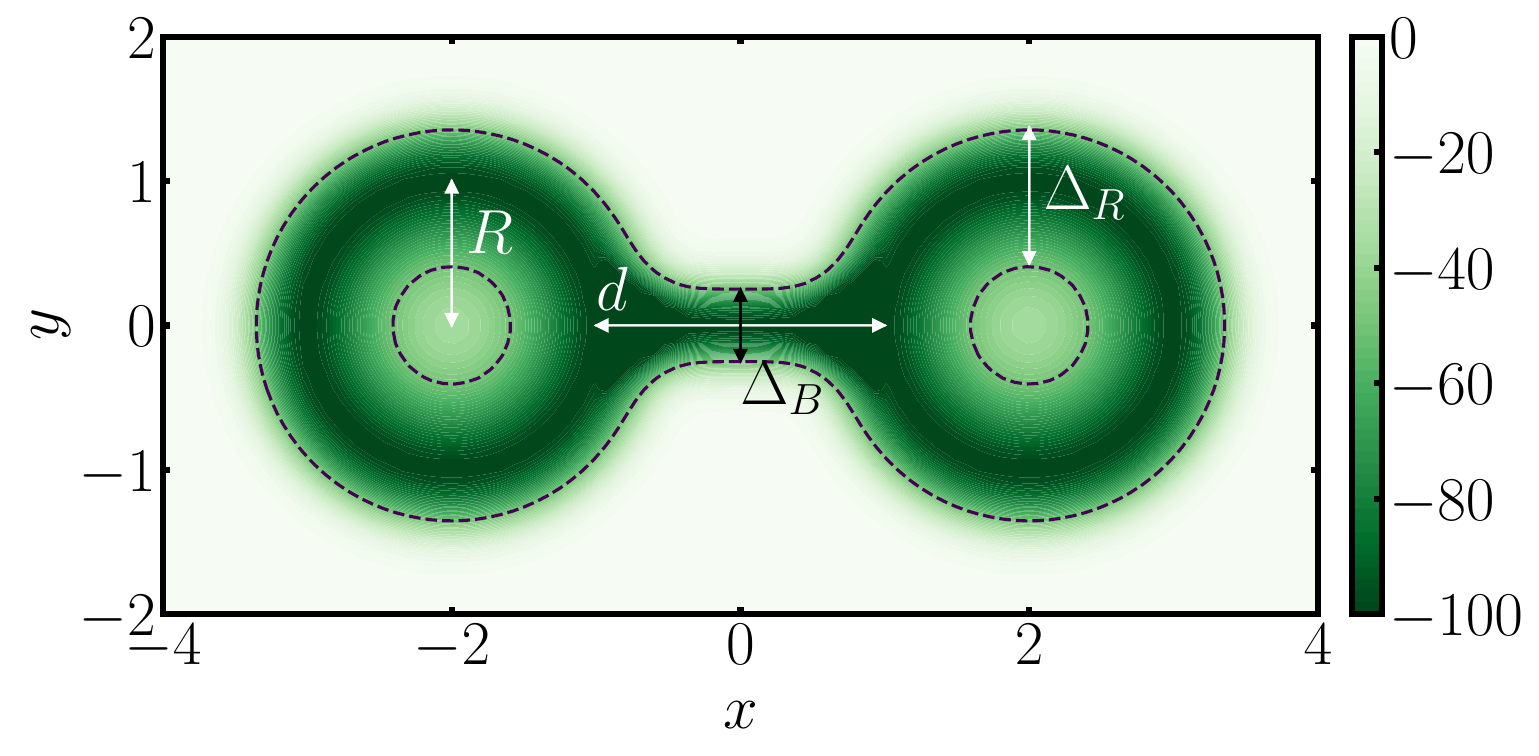}
    \caption{By connecting two Gaussian rings \eqref{eq:Gaussian} with a Gaussian bridge \eqref{eq:Bridge}, we realize the barbell-shaped potential \eqref{eq:Barbell} as the fundamental structure for atomtronic component filtering. The black dashed line shows the contour at ${V_0/2}$. We use harmonic oscillator units.}
    \label{fig:O-O}
\end{figure}

We propose to fix the orientation of the domain interface by designing a barbell-shaped potential that breaks the rotational symmetry. It is composed of two ring-shaped potentials of type \eqref{eq:Gaussian}, centered at ${\vb{r}_\pm = {\pqty{\pm d/2 \pm R, 0}}}$ and connected by a Gaussian bridge
\begin{equation} \label{eq:Bridge}
	V_{B}\pqty{x, y} = - V_0 \;\Exp{- {\pqty{\frac{y}{\Delta_B}}}^2} {\bqty{\eta {\pqty{x+\frac{d}{2}}} - \eta {\pqty{x-\frac{d}{2}}} }} \,,
\end{equation}
with the cartesian coordinates $x$ and $y$ of $\vb{r}$, the radius $R$ of the ring, the length $d$ of the bridge connecting the two rings, and the width $\Delta_B$ of the bridge. We illustrate the resulting barbell potential 
\begin{equation} \label{eq:Barbell}
    \begin{aligned}
		V\rewop{ext} {\pqty{x, y}} ={}& \max \left\{-V_0\,,\;V_{B} {\pqty{x, y}}+\sum_{s=\pm} V_{R} {\pqty{\vb{r} - \vb{r}_s}}\right\}
    \end{aligned}
\end{equation}
in \Fig{O-O}, where `max' ensures flat interstitial regions between bridge and rings. Unless stated otherwise, we use the default parameters given in Table~\ref{tab:default}.

\begin{table}[htb!]
    \centering
    \begin{tabular}{c | c c c c c c c}
        \hline \hline
        Parameters & $N^{\pm}$ & $c$ & $R$ & $d$ & $\Delta_R$ & $\Delta_B$ & $V_0$ \\
        \hline
        Values & 10 & 8.0 & 1.0 & 2.0 & 2.0 & 0.3 & 100.0 \\
        \hline \hline
    \end{tabular}
    \caption{Default parameter values for simulations with the barbell potential \eqref{eq:Barbell}, see also \Fig{O-O}.}
    \label{tab:default}
\end{table}

\section{Results} \label{sec:Results}

In the following, we document our results from designing an atomtronic element for the controlled filtering of Fermi gas components under realistic experimental conditions.

\subsection{Few particles with bare contact interaction} \label{sec:10+10}

\begin{figure*}[htb!]
	\centering
	\begin{picture}(200,240)
		\put(-155,-5){\includegraphics[width=0.55\linewidth]{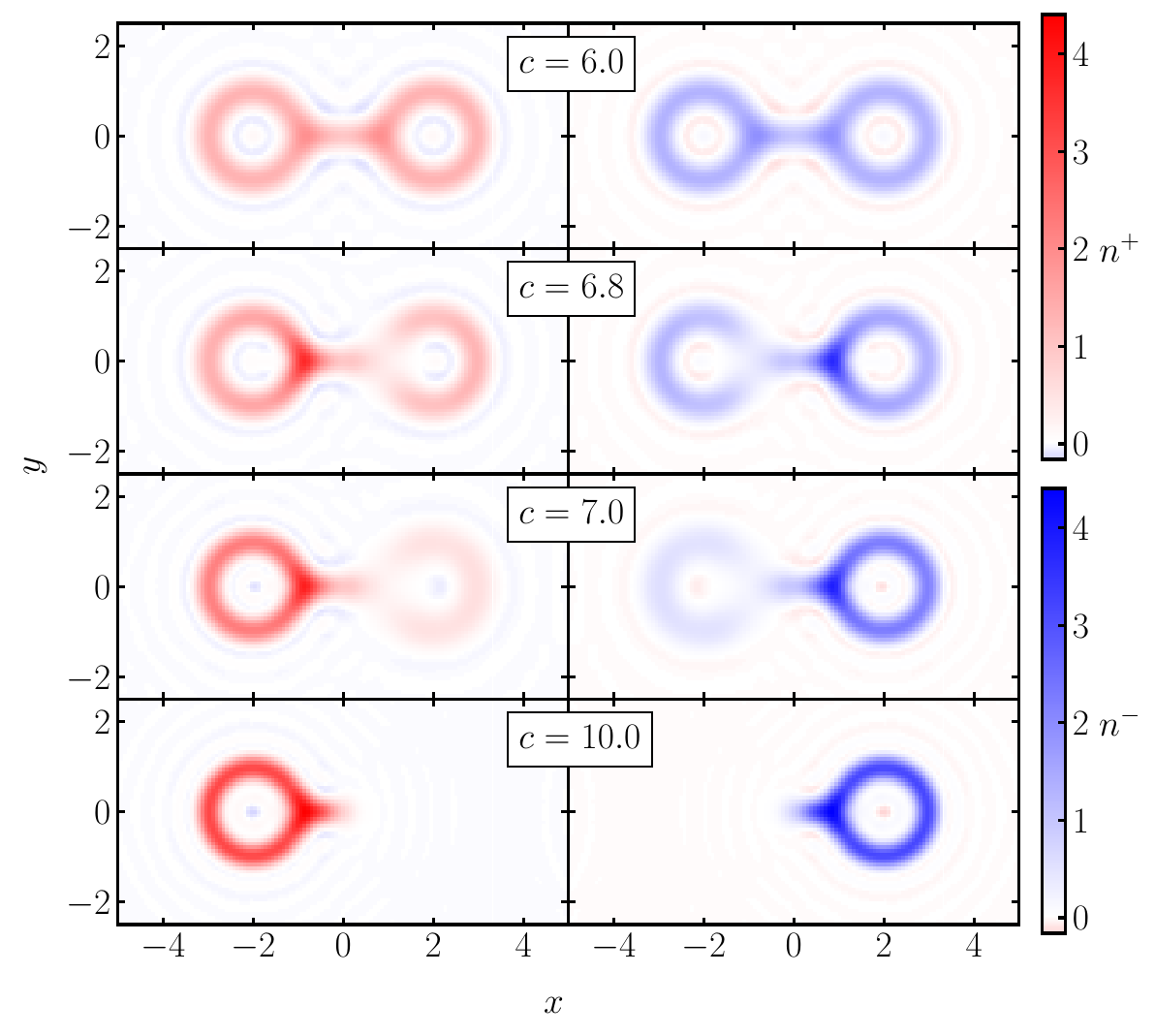}}
		\put(120,1){\includegraphics[height=62.5ex]{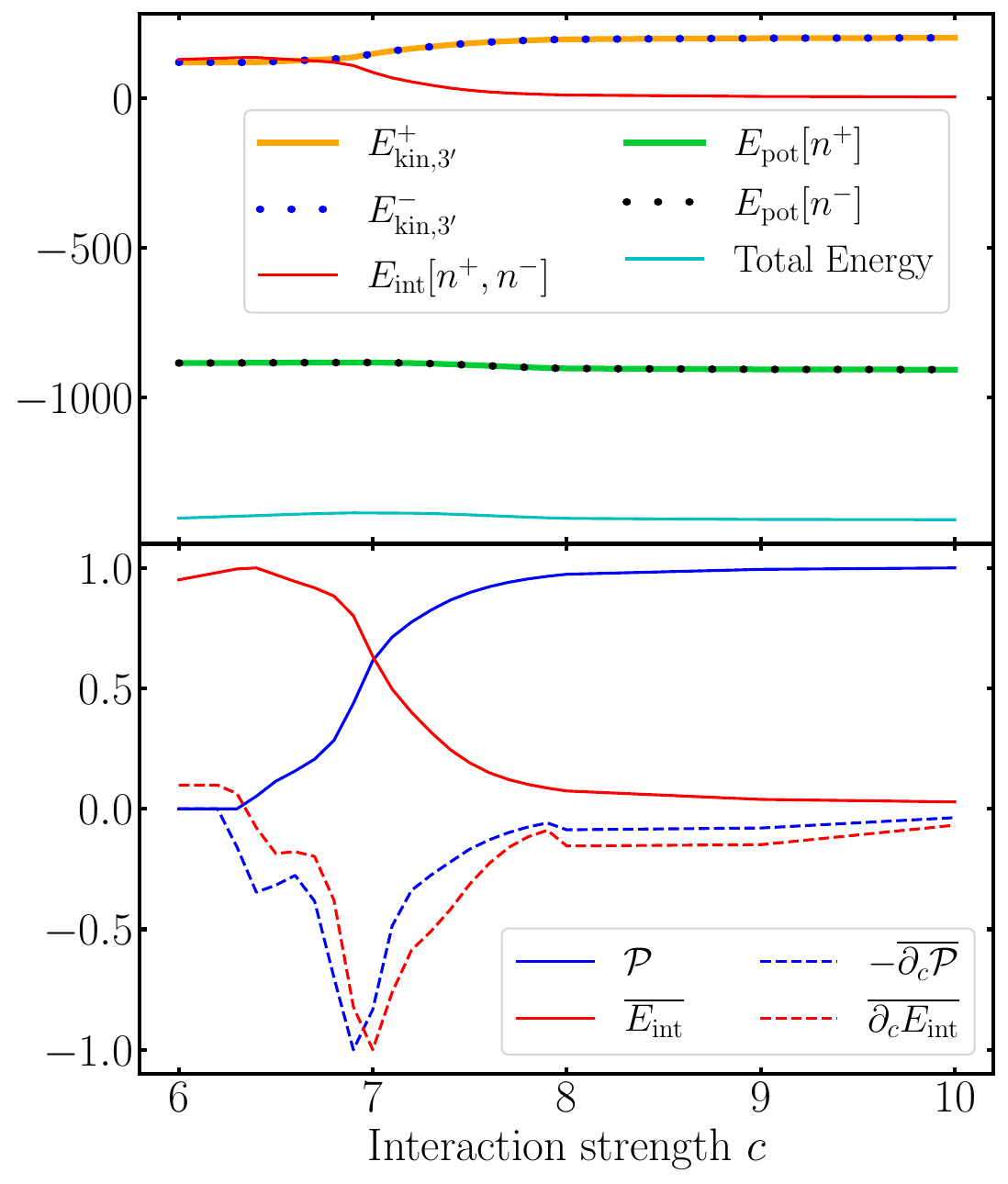}}
		\put(-155,235){(a)}
		\put(120,235){(b)}
		\put(120,125){(c)}
	\end{picture}
	\caption{Transition of a two-component Fermi gas in a barbell-shaped potential from a mixed phase into a filtered, split phase. (a) The sweep of repulsion $c$ from 6.0 to 10.0 (see Table~\ref{tab:default} for the other parameters) reveals the increasing spatial separation between the Fermi gas components, which respect the mirror symmetry of the trapping potential. (b) The phase transition is accompanied by a redistribution between the kinetic energy of both components and the interaction energy, while the potential energies and the total energy show comparably little variation, because the density redistributions occur preferably at small slopes of the external potential, i.e., at the bottom of the trap. (c) The trade-off between kinetic and interaction energy can be visualized with the polarization \eqref{eq:Polarization}, which anti-correlates with the interaction energy; we rescale quantities $Q$ for presentational purposes according to ${\overline{Q} = Q/\max \abs{Q}}$. With this normalization, we also find the gradients $-\overline{\partial_c \mathcal{P}}$ and $\overline{\partial_c \EE{int}}$ closely correlated. We use harmonic oscillator units.}
	\label{fig:Standard}
\end{figure*}
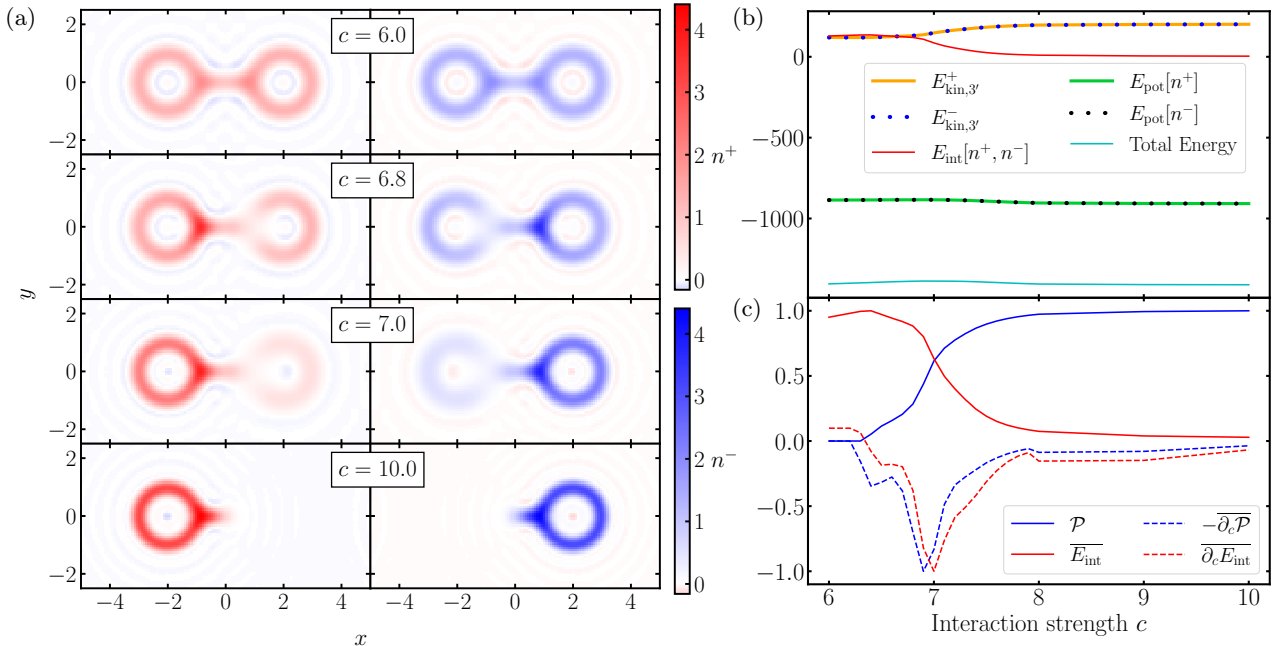

First, we consider small particle numbers, where the quantum effects beyond the TF approximation are relatively more pronounced and need to be captured with quantum-corrected density expressions like $n_{3'}$ in \eqref{eq:n3'}. In contrast to the TF approximation, this nonlocal quantum-corrected density allows us to map the phase transition unambiguously. Figure \ref{fig:Standard}(a) shows how the two Fermi gas densities $n^+$ and $n^-$ with particle numbers ${N^{\pm} = 10}$ in the barbell potential \eqref{eq:Barbell} separate spatially with increasing bare contact interaction strength $c$. At ${c=10}$, the transition is essentially complete, and the components are stored in one ring each.

The kinetic and external energies of both components in \Fig{Standard}(b) are the same across the repulsion sweep because their density profiles do not break the mirror symmetry of the system, see \Fig{Standard}(a). The cause of the phase transition toward complete separation is the competition between the kinetic energies and the interaction energy, as depicted in \Fig{Standard}(b), in line with the qualitative behavior predicted by the TF model \cite{Trappe2021b}. As an alternative for quantifying the phase transition, we plot in \Fig{Standard}(c) the interaction energy alongside the polarization 
\begin{equation} \label{eq:Polarization}
	\mathcal{P} = \frac{1}{N^++N^-} \int {\pqty{\dd{\vb{r}}}} \abs{n^+ {\pqty{\vb{r}}} - n^- {\pqty{\vb{r}}}}\,,
\end{equation}
which measures the overlap of both components, where ${\mathcal{P} = 0}$ indicates the completely mixed phase (${n^+ {\pqty{\vb{r}}} = n^- {\pqty{\vb{r}}}}$ everywhere) and ${\mathcal{P} = 1}$ the full split.

\begin{figure*}[t]
	\centering
	\begin{picture}(200,220)
        \put(-160,0){\includegraphics[width=\linewidth]{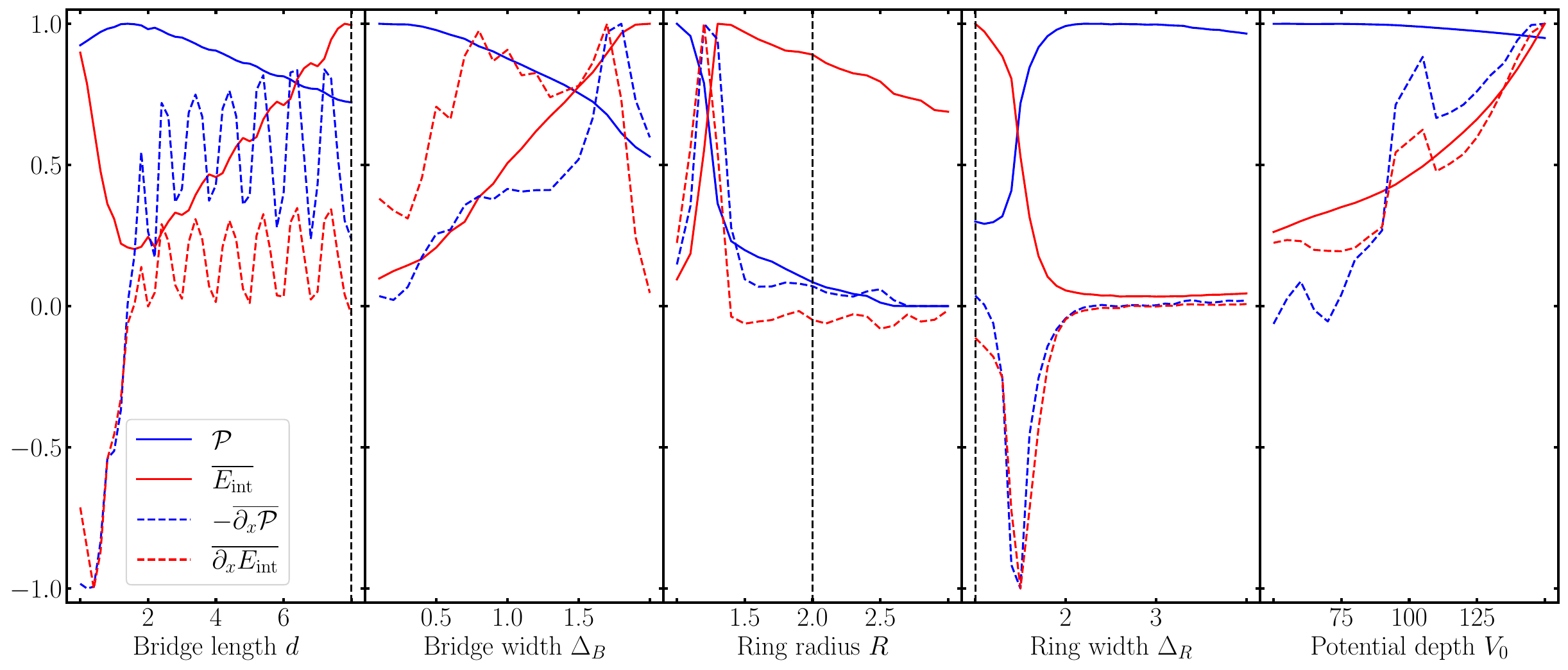}}
		\put(-60,35){(a)}
		\put(37,35){(b)}
		\put(134,35){(c)}
		\put(231,35){(d)}
		\put(328,35){(e)}
	\end{picture}
	\caption{Exploration of the parameter space of the barbell potential \eqref{eq:Barbell}. Each panel shows polarization $\mathcal{P}$, interaction energy $\EE{int}$ ($\overline{Q}$ denotes ${Q/\max \abs{Q}}$), and their derivatives $\partial_x$ as one of the parameters (here denoted as $x$) from Table~\ref{tab:default} is varied: bridge length $d$ (a), bridge width $\Delta_B$ (b), ring radius $R$ (c), ring width $\Delta_R$ (d), and depth $V_0$ of the potential energy (e). We find correlation patterns between $\mathcal{P}$ and $\EE{int}$ that are consistent with those in \Fig{Standard} and, importantly, persist as all parameters are varied individually. The vertical black dashed lines mark the parameters of the special cases shown in \Fig{Others} below. We use harmonic oscillator units.}
	\label{fig:OthersPol}
\end{figure*}

Next, we investigate how the parameters of the barbell potential in \eqref{eq:Barbell} affect the transition toward the fully separated phase of the filter component. For that purpose, we plot the polarization and the interaction energy in \Fig{OthersPol} as functions of the individual barbell parameters, while keeping ${c = 8.0}$ fixed as the reference interaction strength at which the system is already well separated, see \Fig{Standard}.

As suggested by \Fig{OthersPol}(a), (b), and (c), we can \emph{reduce} the polarization significantly by increasing bridge length, bridge width, and ring radius: The Fermi gas transits from a fully separated phase into a phase that has both components $n^+$ and $n^-$ partially mixed.
One might be tempted to trace this decay of polarization to the increase of `trap volume' (roughly, trapping area $\times$ potential depth). In the following we discuss to which extent this narrative holds.

In general, polarisation of the Fermi gas arises from the competition between the kinetic and the interaction energy once the latter exceeds a critical value. Local separations, i.e., local deviations from $n^\pm\rewop{(mix)}$, can be observed first in regions with small gradients $\nabla V$, where density redistributions hardly incur costs of external energy. Then, upon expansion of regions with small $\nabla V$ at fixed $c$, a larger area becomes available for polarization---with accordingly smaller amplitudes of density variations around $n\rewop{(mix)}$. One might argue that these variations should be small enough to imply smaller polarization or void their energetic advantage altogether. Conversely, given the onset of separations at the bottom of the trap, one might argue that regions of small $\nabla V$ facilitate---and, hence, actually increase---polarization.

Both these quasiclassical narratives are at odds with the numerical evidence: First, the trend of diminishing polarization with increasing `trap volume' is broken in the case of the ring width $\Delta_R$, see \Fig{OthersPol}(d), where the system gradually separates and then largely remains in a fully separated phase with increasing $\Delta_R$. This somewhat counter-intuitive example reminds us of the importance to augment the quasiclassical picture (based in particular on the intuitively accessible TF model) with systematic quantum corrections such as those provided within the DPFT framework. Second, in similar numerical experiments with fermions in harmonic traps, we find that varying trap frequency neither yields a monotonic upward nor a monotonic downward trend in total polarization.

In summary, the polarization curves from our \emph{quantum-corrected} DPFT simulations in \Fig{OthersPol} suggest that the barbell potential will have a strong filtering capacity if we choose a short and narrow bridge, a small ring radius, and a large ring width. The potential depth is less important, as long as the trap is deep enough to accommodate the prescribed particle number, here ${N^{\pm} = 10}$, see \Fig{OthersPol}(e). We emphasize, however, that these quantitative findings depend on the particle number and interaction kernels. That is, \Fig{OthersPol} informs about the efficacy of parameter configurations around the default parameters of Table~\ref{tab:default}. Hence, our work provides the procedural blueprint for reassessing vastly different parameter regimes.

\begin{figure}[ht!]
    \centering
    \includegraphics[width = \linewidth]{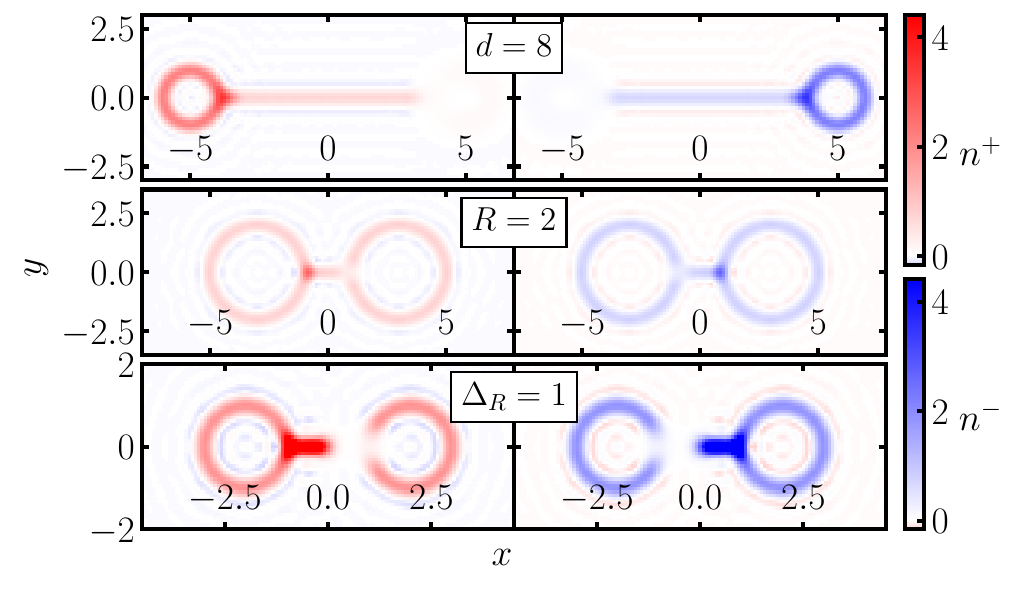}
    \put(-210,130){(a)}
    \put(-210,87){(b)}
    \put(-210,45){(c)}
	\caption{Partially mixed density profiles in various parameter regimes of the barbell potential, cf. the black dashed lines in \Fig{OthersPol}. The changes relative to Table~\ref{tab:default} are given in the boxes. We use harmonic oscillator units.}
    \label{fig:Others}
\end{figure}

In \Fig{Others} we present additional density profiles that differ qualitatively from the partial separations shown in \Fig{Standard} and provide further understanding of the functioning of the spin filter: Depending on which barbell parameters are modified, the Fermi gas components can be equally mixed in the bridge while they are fully separated in the rings (\Fig{Standard}, top), or vice versa (\Fig{Standard}, bottom), or in between both extremes (\Fig{Standard}, centre).

\subsection{Many particles under realistic conditions} \label{sec:realistic}

Guided by the results in the previous \Sec{10+10}, we shall now test the barbell potential under laboratory-like conditions. First, apart from the parameters of the barbell potential, also the functional form of the repulsive contact interaction influences the phase transitions of the two-component Fermi gas \cite{Trappe2021b}. However, when replacing the bare contact interaction in \eqref{eq:Contact} by the more realistic, dressed contact interaction in \eqref{eq:Dressed}, we observe the same trend of component separation: Figure \ref{fig:Dressed} in the appendix for ${N^{\pm} = 10}$ particles mirrors \Fig{Standard}, suggesting that the bare contact interaction is sufficient for designing the atomtronic spin filter. In the following we simulate experimentally relevant particle numbers with DPFT and further analyse the impact of both interaction kernels and kinetic energy functionals.

\begin{figure}[ht!]
    \includegraphics[width = \linewidth]{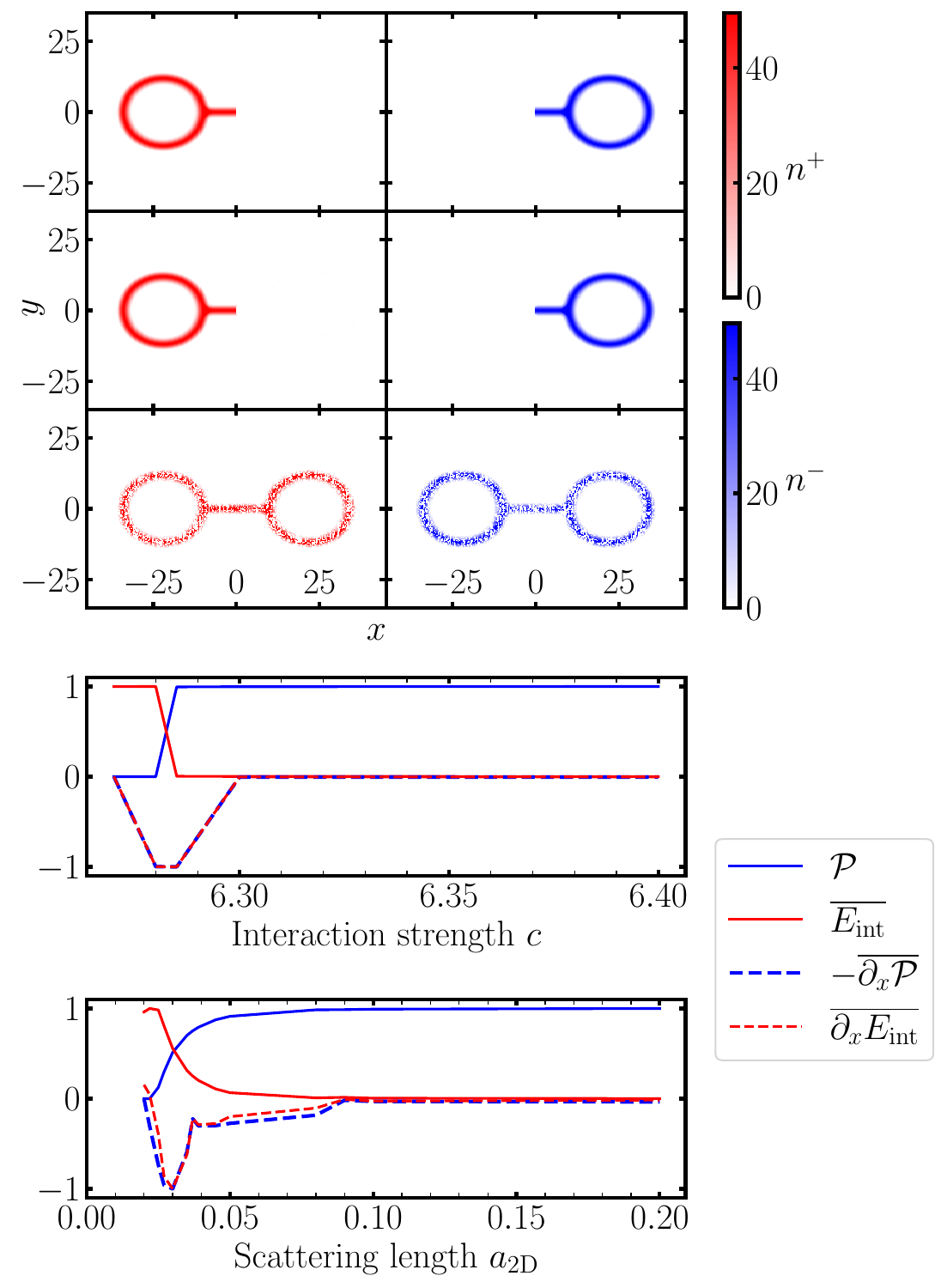}
    \put(-220,320){(a)} 
    \put(-220,270){(b)}
    \put(-220,217){(c)}
    \put(-220,145){(d)}
    \put(-223,60){(e)}
	\caption{Realistic spin filtering via the here developed barbell potential. Our parameter choices align with the experimental setting reported in \cite{Cai_2022} for ${N^{\pm} = 10000}$ Lithium-6 atoms. The completely separated density profiles for the dressed contact interaction at ${a_{\mathrm{2D}}=0.2}$ in panel (a) are indistinguishable (to the eye) from the profiles based on the bare contact interaction at ${c=6.285}$ in panel (b). Panel (c) shows TF density profiles at the same ${c=6.285}$, where the random separation patterns originate from numerical noise that gets amplified due to the spatial decoupling in the (consequently inadequate) TF model. Panel (d) identifies the phase transition towards complete separation at the critical interaction strength ${c_{\mathrm{split}}\approx 2\pi}$ for the bare contact interaction, and panel (e) identifies a transition window in the vicinity of ${a_{\mathrm{2D}}\approx 0.03}$ for the dressed contact interaction. We use harmonic oscillator units.}
    \label{fig:Real}
\end{figure}
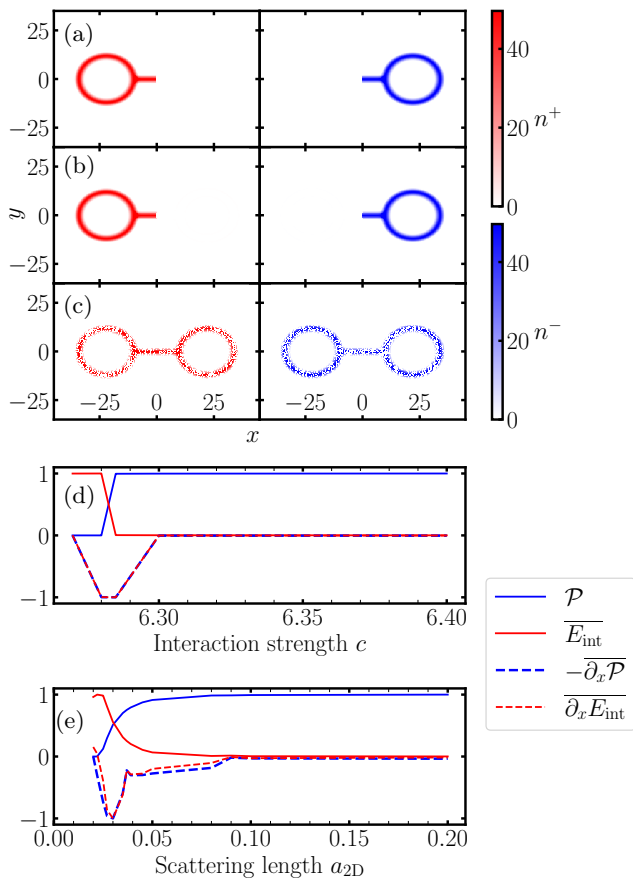

To demonstrate the functioning of our component filter under realistic conditions, we borrowed the experimental parameters reported in \cite{Cai_2022}, where approximately $20000$ Lithium-6 atoms are trapped in a ring-shaped potential. Accordingly, we simulated ${N^{\pm} = 10000}$ particles in the barbell potential with the parameters ${R = 12 \mathrm{\mu m}}$, ${d = 20 \mathrm{\mu m}}$, ${\Delta_R = 12.1165 \mathrm{\mu m}}$, ${\Delta_B = 1.5 \mathrm{\mu m}}$, and ${V_0 = 350}$. The results are shown in \Fig{Real}.

As expected, the two components become fully separated for strong enough repulsion. This is the case for both interaction kernels, see \Fig{Real}(a) and \Fig{Real}(b). Moreover, as the number of particles increases, the phase transition toward a full split sharpens (compare \Fig{Real}(d) with \Fig{Standard}(c)) in the case of the bare contact interaction, where the transition occurs between ${c=6.28}$ and ${c=6.285}$, i.e., not unexpectedly, closer to the TF value ${c_{\mathrm{split}}^{\mathrm{TF}}=2\pi}$, see \cite{Trappe2021b}, when compared with the few-particle situation in \Fig{Standard}. Concerning the TF model we note, however, that (for the purely local contact interactions) the TF approximation decouples different spatial positions in the self-consistent equations \eqref{eq:SelfConsist}, such that numerical rounding-off errors induce random patterns of fine-grained separations, see \Fig{Real}(c)---with unrealistically high (exact) kinetic energy---instead of the actual bipartite separation \cite{Trappe2021b}. Hence, convergence to the ground-state densities requires ad-hoc manual interventions or, preferably, nonlocal quantum-corrected density expressions like $n_{3'}$. It is important to note that the bare contact interaction yields a sharp phase transition compared to the dressed contact interaction, which induces a smooth crossover, with \Fig{Real}(e) pinpointing the midpoint of the transition (viz.~$\mathcal P=0.5$) to $a_{\mathrm{2D}}\approx0.03$. This gradual rather than sharp transition at large particle numbers is the primary difference between both interaction kernels and should be falsifiable with present experimental techniques.

\section{Conclusions} \label{sec:Conclusions}
With the help of density potential functional theory (DPFT), we investigated the behavior of a two-component Fermi gas with repulsive contact interaction in a ring-shaped potential. From two such rings we designed a barbell-shaped potential that can spatially separate the two components via a connecting bridge if the repulsive interaction is strong enough. In other words, we proposed a filter for fermionic (spin-)components and thus added a new circuit element to the field of atomtronics, complementing (bosonic) diode/transistor functionalities in multi-well devices \cite{haug2019aharonov,Pepino2009_PRL,Thorn2008,Seaman2007_PRA}. We also characterized the phase transitions that can be observed in this filter across interaction strengths, and we investigated the separation profiles of the fermionic clouds as a function of the barbell potential parameters. We concluded our comprehensive characterization by demonstrating that the barbell component filter works in the parameter regime of recent atomtronics experiments.

Our results provide a natural route to state- or mixture-selective transport in two-component Fermi gases via interaction-controlled phase separation along the bridge of the barbell potential. The functionality of this \emph{passive} component filter is rooted in fundamental many-body interactions and a suitably fixed (rather that dynamically manipulated) trap geometry. It therefore lends itself as a robust component of integrated atomtronic circuits. In essence, our results establish (i) the barbell potential as a robust component for passive spin filtering and (ii) showcase the efficacy of the DPFT framework to design and validate atomtronic devices under realistic conditions relevant to current experiments with mesoscopic Fermi gases. The design of new atomtronics elements relevant to experimenters is merely one capability of DPFT. Although this orbital-free DFT variant excels especially in simulating interacting Fermi gases with large particle numbers across dimensions, DPFT is a powerful method for simulating many-body quantum systems in general: incorporating finite temperature, noise, and other experimental constraints is straightforward.

We therefore argue that the DPFT framework can serve as a simulator of integrated atomtronic setups that mirror the components of traditional electronic or optical circuits, such as filters, polarizing beam splitters, switches, capacitors and memory, or logic gates. For example, the outputs of the barbell filter could be guided into matter-wave interferometers to explore spin-sensitive quantum transport phenomena or could act as two parallel sources feeding a shared spin--orbit gate of a fermionic Datta-Das transistor \cite{vaishnav2008spin}. We also note that beyond passive action, temporal shaping \cite{grochowski2025quantum} of the barbell potential might accelerate separation dynamics. Last but not least, we mention two other possible applications of our fermionic separator. A fundamental mechanism of batteries lies in manipulating the chemical potentials. In recent years, there has been a keen interest in developing quantum batteries \cite{campaioli2024colloquium}.  
It is conceivable for two species of fermions to provide the different chemical potentials needed for a battery \cite{rossini2020quantum,wang2012optimization,konar2022quantum}. A fermionic filter would be a handy device for reversing the mixing of a two-species battery.  As another example, we consider the quantum heat engine\cite{arezzo2024many,chen2019interaction}. A quantum many-body engine fueled by the energy difference between fermionic and bosonic ensembles of ultracold particles\cite{koch2023quantum} has already been realized in experiment. Incorporating a quantum filter into such devices may enhance the efficiency of the heat engine.

\acknowledgements

This project is supported by the National Research Foundation, Singapore, through the National Quantum Office, hosted in A*STAR, under its Center for Quantum Technologies Funding Initiative (Grant No. S24Q2d0009). The authors also acknowledge CQT funding under the National Research Foundation and the Ministry of Education. P.T.G. was supported by the project CZ.02.01.01/00/22\_010/0013054 (C-MONS) within Programme JAC MSCA Fellowships at Palack\'y University Olomouc IV.

\section*{Appendix}

\begin{figure}[ht!]
    \centering
    \includegraphics[width = \linewidth]{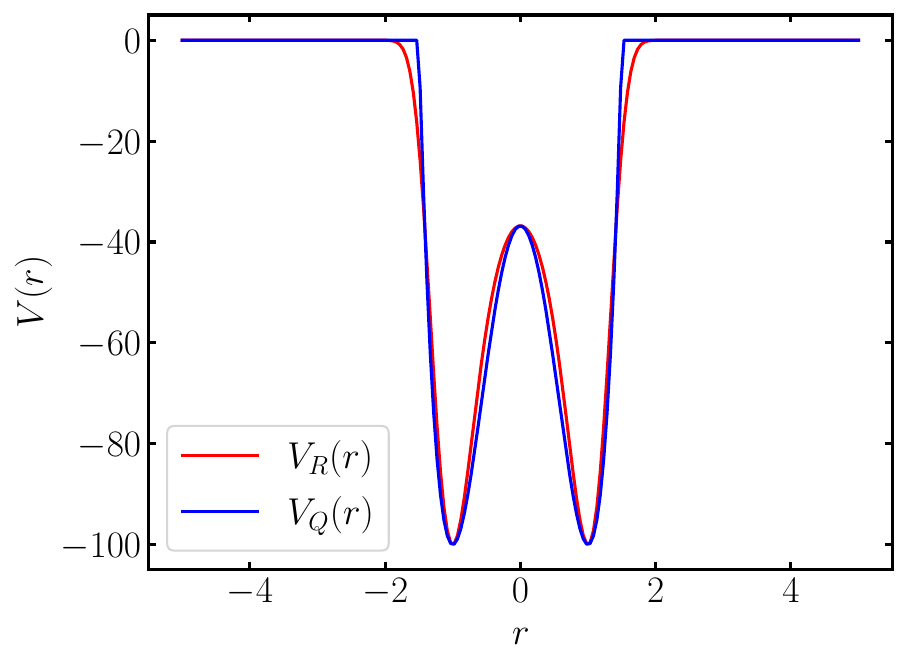}
	\caption{The radial plot for the two choices of ring trapping potential, where the maxima and minima of the Gaussian ring (red) matches those of the quartic ring (blue). Here, the quartic ring potential is cut off when it is larger than $0$, and parameters are those of Table~\ref{tab:default}. The minute differences between $V_R$ and $V_Q$ are intentional and a proxy for the imperfect replication of the trapping potential in experiments. We use harmonic oscillator units.}
    \label{fig:RingPotentials}
\end{figure}

In \Sec{Filter} and throughout the main text, we employed the Gaussian-shaped ring potential $V_R {\pqty{\vb{r}}}$ in \eqref{eq:Gaussian}. In a practical experiment, however, the prescribed shape of a trapping potential can only be created approximately---ideally, therefore, theoretical designs are robust against small deviations from the intended trap geometry. To investigate the extent to which such variations in the experimental design affect the separation of the two components, we compare density profiles based on $V_R {\pqty{\vb{r}}}$ and the alternative quartic potential
\begin{equation} \label{eq:Quartic}
	V_Q \pqty{\vb{r}} = D {\pqty{r^4 - 2 R^2 r^2}} + B \,,
\end{equation} 
where $D$ controls the depth of the potential, $B$ is the bias of the potential, and $R$ controls the radius of the ring. We align the two potentials $V_Q {\pqty{\vb{r}}}$ and $V_R {\pqty{\vb{r}}}$ via
\begin{equation}
    B = - V_0 \Exp{- \frac{16}{\Delta_R^4} R^4} \qand D = \frac{B + V_0}{R^4} \,,
\end{equation}
see \Fig{RingPotentials}.

When repeating the simulations for \Fig{Ring} with $V_Q$ as external potential, we observe merely slight differences in the critical interaction strengths (compared with the case of $V_R$) at which the number of domains changes (with differences on the order of ${\Delta c\approx0.01}$), for example, ${c_{\mathrm{split}}(V_Q)\approx 6.30}$, compared with ${c_{\mathrm{split}}(V_R)\approx6.31}$. Other than that, the entire sequence of density profiles shown in \Fig{Ring}, from a symmetrically mixed phase to the distinct domain separations to a symmetrically split phase, is quasi indistinguishable to the case with $V_Q$, which also underscores the robust design of our component filter. Other experimental constraints like moderate noise, temperature, and slight deviations from a purely two-dimensional trap geometry can also be expected to not change the qualitative picture \cite{Trappe2021b,Grochowski2024}.

\begin{figure}[ht!]
    \centering
    \includegraphics[width=\linewidth]{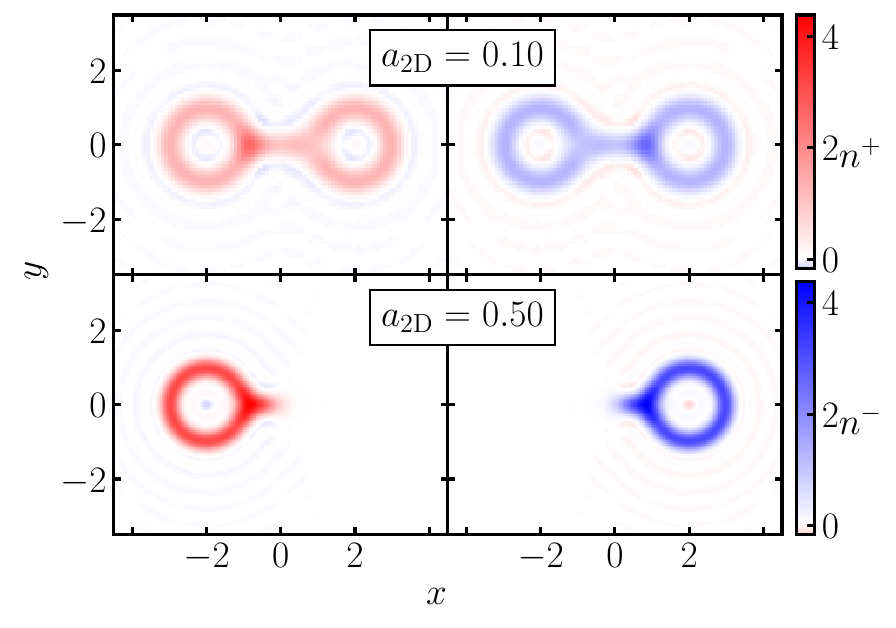}
	\caption{Density profiles as in \Fig{Standard}(a), with default parameters from Table~\ref{tab:default}, but with the dressed contact interaction \eqref{eq:Dressed}: The transition toward full separation is qualitatively similar when compared with the bare contact interaction \eqref{eq:Contact}. We use harmonic oscillator units.}
    \label{fig:Dressed}
\end{figure}

\bibliographystyle{apsrev4-2}
\bibliography{Hue_et_al_Atomtronics}

\end{document}